\documentclass{emulateapj}
\usepackage{apjfonts}

\newcommand{\Chandra}{{\em Chandra}}
\newcommand{\HST}{{\em HST}}
\newcommand{\BW}{{\em BW}}
\newcommand{\SW}{{\em SW}}

\begin{document}

\title{{\em ChaMPlane} Discovery of Candidate Symbiotic Binaries in
Baade's and Stanek's Windows}

\slugcomment{To appear in ApJ Letters}

\author{M. van den Berg\altaffilmark{1}, J.  Grindlay\altaffilmark{1},
S. Laycock\altaffilmark{1}, J.  Hong\altaffilmark{1},
P. Zhao\altaffilmark{1}, X.  Koenig\altaffilmark{1},
E. M. Schlegel\altaffilmark{2}, H.  Cohn\altaffilmark{3},
P. Lugger\altaffilmark{3}, R. M. Rich\altaffilmark{4}, A. K.
Dupree\altaffilmark{1}, G. H.  Smith\altaffilmark{5},
J. Strader\altaffilmark{5}} \altaffiltext{1}{Harvard-Smithsonian
Center for Astrophysics, 60 Garden Street, Cambridge, MA 02138, USA;
maureen@head.cfa.harvard.edu} \altaffiltext{2}{Univ. of Texas, Dep. of
Physics and Astronomy, San Antonio, TX 78249} \altaffiltext{3}{Indiana
Univ., Dep. of Astronomy, Bloomington, IN 47405}
\altaffiltext{4}{UCLA, Dep. of Physics and Astronomy, Los Angeles CA,
90095-1547} \altaffiltext{5}{Univ. of California Observatories/Lick
Observatory, Dep. of Astronomy and Astrophysics, Santa Cruz, CA 95064}

\begin{abstract}
We have searched the {\em OGLE-II} archive for candidate counterparts
of X-ray sources detected in two low-extinction windows included in
our Galactic bulge {\em Chandra/HST} survey. We find that a
significant number---i.e.~in excess of the expected level of random
associations---can be matched with probable M-giants.  Their X-ray
properties can be understood if these sources are symbiotic binaries
where the X-rays are typically, either directly or indirectly, the
result of a white dwarf accreting from the wind of a cool
giant. Optical and near-infrared properties of selected sources are
consistent with a symbiotic nature, although none of the spectra
collected for 8 out of 13 candidate counterparts show the
high-ionization nebular emission lines observed for many
symbiotics. The hard X-ray emission for several sources (power-law
photon indices $-1.5\lesssim\Gamma\lesssim1.5$) suggests our sample
includes systems similar to the symbiotics recently detected with {\em
INTEGRAL} and {\em Swift}.
\end{abstract}

\keywords{X-rays: binaries --- X-rays: stars --- binaries: symbiotic --- Galaxy: bulge --- stars: late-type}

\section{Introduction}

The main goal of the \Chandra~Multiwavelength Plane Survey ({\em
ChaMPlane}\footnote{http://hea-www.harvard.edu/ChaMPlane}) is to
identify and study the populations of low-luminosity
($L_X\lesssim10^{33}$\,erg\,s$^{-1}$) accretion-powered binaries in
the Galaxy \citep{grinhongea05}. {\em ChaMPlane} includes a deep
survey of the Galactic bulge consisting of (near-)simul\-taneous
\Chandra~and {\em Hubble Space Telescope} (\HST) pointings of three
low-extinction regions. With minimal obscuration by dust, for the
bulge, and positions progressively closer to the Galactic Center (GC),
these ``windows'' are chosen to trace the X-ray point-source
population towards the GC. The bulge population of interacting
binaries is also of interest for comparison with those in globular
clusters, where stellar encounters have enhanced the numbers of close
binaries, and with X-ray populations in other galaxies. Here we report
initial results of observations of Baade's Window (\BW,
($l$,$b$)=(1\fdg06,$-$3\fdg83)) and Stanek's Window \citep[\SW,
($l$,$b$)=(0\fdg25,$-$2\fdg15);][]{stan98} where we have found a
significant number of sources with candidate M-giant counterparts. In
$\S$\,\ref{sec_obs} we describe the optical identification, and X-ray
and optical/near-infrared spectra.  As X-ray emission from single
M-giants is rarely detected \citep{huenea98}, we have investigated
whether these sources could be symbiotic binaries in which a cool
giant typically transfers mass to a white dwarf via a wind
($\S$\,\ref{sec_dis}). More results, including analysis of our
``Limiting Window'' at ($l$,$b$)=(0\fdg10,$-$1\fdg43) will be
presented in future papers.

\section{Data and Analysis} \label{sec_obs}

\subsection{Chandra Data} \label{ssec_xray}

We obtained \Chandra/ACIS-I observations of \BW~(obsid\,3780; 98 ks)
and \SW~on 2003 July 9 and 2004 February 14/15, respectively. The
{\SW} observation was split in two exposures of 80 ks (obsid\,4547)
and 19 ks (obsid\,5303), with a gap of 1\fh2, to keep the same roll
angle; these were stacked for the remainder of the analysis.  The {\em
ChaMPlane} X-ray data reduction pipeline \citep{hongvandea05} was used
to create source lists from images in the 0.3--8 keV band and 95\%
confidence radii on positions $r_{95\%}$. A total of 365 (\BW) and 389
(\SW) ACIS-I sources were detected.  Other pipeline products include
net source counts between 0.5--8 keV, and energy quantiles $E_x$
corresponding to the energies below which $x$\% of the counts are
detected.

\subsection{Optical Identification with OGLE-II Sources} \label{sec_oid}

\begin{figure}
\centerline{\includegraphics{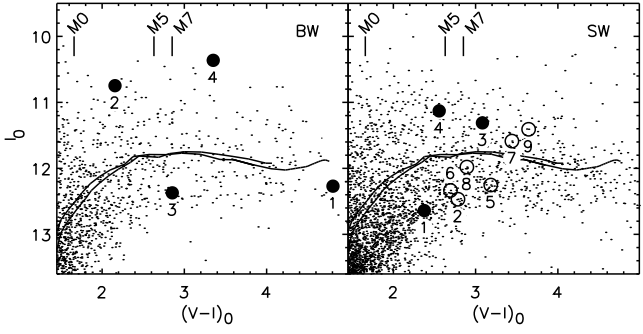}}
\caption{Color-magnitude diagrams for \BW~and \SW~(U02), corrected for
the typical ($E$($V$$-$$I$), $A_I$)$\approx$(0.84,0.81) in \BW~and
(1.32,1.27) in \SW. Matches from Table~\ref{tab_prop} are labeled; an
open symbol indicates there are multiple U02-stars in the 2\,$\sigma$
match circle. The curves are the first and asymptotic giant branches
of a $Z$=0.019, 10-Gyr isochrone \citep{gira06} at 8 kpc; the spread
of stars around the isochrone can be due to distance or non-solar
compositions. Expected colors for M-giants are marked
\citep{houdbellea00}.
\label{fig_cmd}}
\end{figure}

We searched the Optical Gravitational Lensing Experiment II ({\em
OGLE-II}) star catalog \citep[U02 hereafter]{udalszymea02} for
counterparts. Due to the high probability for spurious matches, we
used only variable stars \citep{woznea02} to measure the {\em Chandra}
boresight correction ($\sim$3$\times$10$^{5}$ stars in the ACIS-I
areas, versus $\sim$1$\times$10$^3$ (\BW) and $\sim$4$\times$10$^3$
(\SW) variables). As the U02 and Wozniak et al.\,catalogs are not
cross-linked, we first identified the variables in U02 using match
criteria based on difference in position ($<$2 pixels or
$\sim$0\farcs82\footnote{The catalogs are based on different
photometry methods (direct versus difference imaging); therefore,
stars included in both catalogs need not necessarily be listed with
identical coordinates.}) and $I$ magnitude.  We used the boresight
method described in \cite{zhaogrindea05} adopting a 1\,$\sigma$ error
on optical positions $\sigma_{o}=0$\farcs2\,(U02).  The resulting
\Chandra$-${\em OGLE II} offsets ($\Delta\alpha$, $\Delta\delta$) are
($+$0\farcs05(4),$-$0\farcs05(5)) and
($+$0\farcs22(3),$-$0\farcs40(3)), based on 26 and 49 matches in
\BW~and \SW, respectively. The light curves suggest many of these are
interacting binaries for which we indeed expect enhanced X-rays.
After correcting for boresight, we searched for counterparts in
2\,$\sigma$ radii, with $\sigma$ the qua\-dratic sum of $\sigma_{o}$,
$r_{95\%}$ (normalized to 1$\sigma$) and the boresight error. To
reduce spurious matches we only consider ACIS-I sources with
$r_{95\%}\leq3$\arcsec, yielding 351 and 384 \BW~and \SW~sources, for
which 547 and 684 matches are found for 289 and 313
sources\footnote{including 33 and 73 stars without $V$$-$$I$ color
information}, respectively. Given several initial identifications with
M-giants, bulge giants were selected to be at least as red
\citep[$V$$-$$I\geq1.66$ for solar metallicity;][]{houdbellea00} and
at least as bright ($I\leq13.4$ from 10\,Gyr, $Z=0.019$ isochrones by
\cite{gira06}) as M0\,III giants at the GC (8\,kpc).  The maps by
\cite{sumi04}---based on photometry of bulge clump giants---were used
to correct for reddening and extinction\footnote{For foreground stars
this is incorrect as $E($$V$$-$$I)$ and $A_I$ are
overestimated. Foreground M-giants with $I$$\lesssim$11--12 would
however be saturated in U02.}.  Four (\BW) and 9 (\SW) candidate
counterparts satisfy the criteria (Table~\ref{tab_prop},
Fig.~\ref{fig_cmd}). Repeating the matching for randomly offset X-ray
positions ($\alpha$ and $\delta$ varied with 4\farcs5 increments on a
49$\times$51 grid) shows that in \BW~the average number of M-giant
counterparts found by chance $<$$N_g$$>$, is 1.3$\pm$1.1, and that
$N_g\geq4$ with 4.4\% probability; in \SW~$<$$N_g$$>$=3.3$\pm$1.8, and
$N_g\geq9$ with 0.64\% probability.  Star counts by \cite{zhenea01}
predict $\lesssim$0.4 M-dwarfs in the ACIS-I area that can contaminate
our sample.

The inner 6\arcmin$\times$6\arcmin~of {\em BW} and {\em SW} were
observed with the Advanced Camera for Surveys on \HST.  Optical
identification using these data is the topic of future papers. We note
that for the sources in Table~\ref{tab_prop} in the \HST~pointings
(BW1 and 2) we find that all additional objects in the 2\,$\sigma$
match circles have colors consistent with them being late-type
main-sequence stars at $\gtrsim3$ kpc, which would not normally be
luminous enough for detection \citep[$L_X\lesssim10^{30}$ ergs
s$^{-1}$,][]{schmlief04} in our \Chandra~data.

%SORTED ON DECLINATION
\begin{table} 
\caption{Properties of candidate M-giant counterparts \label{tab_prop}} 
\begin{tabular}{llllllll}
\tableline
\tableline
ID                & Counts & $\Gamma$\tablenotemark{a}/ & $n^{\rm a}_H$ & $\log$$L_{X}$\tablenotemark{a}  & SpT\tablenotemark{b}                        & {\em OGLE-II} ID\tablenotemark{c}         \\
                  &             & $kT$    &   &  (8\,kpc)  &    M$\ldots$                        & \\
\tableline
BW1 &  10(5)       & 0.7    & 2.3     & 31.3(2)&  3\tablenotemark{b}    & 45/113191/1472\\ %XS03780B3\_045    H4 H5
BW2 &  15(5)       & 0.7    & 2.6     & 31.5(2)&  $\ldots$                         & 01/640074 \\ %XS03780B1\_046 
BW3 &  12(5)       & -1.2 & 2.2     & 31.7(2)&  4$-$5\tablenotemark{b}& 45/256236/1224\\ %XS03780B2\_062  F H4 H4
BW4 &  30(7)       & 0.0      & 2.7   & 32.2(1)&  $\ldots$\tablenotemark{b}          & 45/244411  \\ %XS03780B0\_029 
\tableline			        
SW1 & 521(26)      & 1.9 & 4.3        & 33.04(2)&  $\ldots$                            & 39/126098  \\ %XS54547B3\_036 
SW2 &  30(8)       & 2.6 & 4.5        & 31.8(1)&  $\ldots$                        & 04/684358/4865\\ %XS54547B3\_211 
SW3 &  54(9)       & 1.5$\tablenotemark{a}$ & 60$\tablenotemark{a}$  &32.88(7)&  5$-$6                 & 04/293339/4625\\ %XS54547B3\_267 F H4
SW4 &  19(6)       & 1.1 & 3.8        & 31.6(1)&  3                        & 04/293323/4310\\ %XS54547B3\_277 F
SW5 &  16(8)       & 1.5 & 4.1        & 31.5(2)&  $\ldots$                         & 04/065565/3277\\ %XS54547B3\_375 
SW6 &  18(9)       & 1.2 & 4.2        & 31.6(2)&  4$-$5                    & 04/065539/3113\\ %XS54547B3\_372 F H4
SW7 & 124(13)      & 0.6 & 4.1        & 32.48(4)&  3$-$5                        & 04/266999/2810\\ %XS54547B3\_164 F
SW8 &  10(6)       & 1.0 & 4.1        & 31.3(3)&  4$-$6                    & 04/648022/2915\\ %XS54547B3\_218   H4
SW9 &  23(10)      & 1.0 & 4.5        & 31.8(2)&  4$-$6                    & 04/051718/2234\\ %XS54547B3\_363 F H4
\tableline
\end{tabular}
\tablenotetext{a}{X-ray properties refer to the 0.5--8 keV
band. Power-law photon index $\Gamma$ as derived from quantile
power-law grids by fixing $n_H$ (given in units of 10$^{21}$
cm$^{-2}$) to the corresponding optically derived $A_V$ for bulge
stars \citep{sumi04} except for SW3 for which $n_H$ and $kT$ are
derived from a thermal-bremsstrahlung grid ({\em see text}). For
$kT$=1~keV thermal bremsstrahlung, 10$^{-4}$ ct s$^{-1}$ correspond to
an unabsorbed flux of 1.36$\times$10$^{-15}$ (\BW;
$n_H$=2.5$\times$10$^{21}$ cm$^{-2}$) and 1.69$\times$10$^{-15}$ ergs
cm$^{-2}$ s$^{-1}$ (\SW; $n_H$=4.2$\times$10$^{21}$ cm$^{-2}$). To
compute $L_X$ (erg s$^{-1}$) a distance of 8\,kpc is assumed.}
\tablenotetext{b}{SpT=spectral type. \cite{blanmccaea84} classify BW1
(their star 86) as M9, BW4 (star 93) as M6; \cite{blan86} give M5 for
BW3 (star 83). A nearby bright star contaminates our spectra of BW1.}
\tablenotetext{c}{IDs given as [field number]/[U02 ID]/
[\cite{woznea02} ID].}\end{table}

\subsection{X-ray Spectral Properties} \label{sec_xspec}

Since most of the selected sources have only 10--30 counts, we use
quantile analysis \citep{hongschlea04} to derive spectral
information. Their median energies $E_{50}$ are relatively high with
all \BW~and 5 \SW~sources having $E_{50}\geq2.5$ keV. Of all the
detected sources in \BW~and \SW, only 14\% (\BW) and 20\% (\SW) have
$E_{50}\geq2.5$ keV.  Spectra for about half of the sources can be
described with a bremsstrahlung model ($kT\gtrsim1.5$ keV;
Fig.~2a). The remaining sources are too hard; power-law models
constrain their photon indices to $-1.5\lesssim\Gamma\lesssim1.5$
(Fig.~2b). Adding a broad 6.4 keV emission line brings these points
even closer to the grid and increases $\Gamma$ (Fig.~2c); this suggest
the presence of a fluorescent Fe\,K line to avoid the unphysical
$\Gamma\lesssim0$ in the hardest sources. The column density $n_H$ is
consistent with the optically-derived value for bulge stars in \BW~and
\SW. An exception is SW3 for which the inferred $n_H$ exceeds the
field value by a factor of $\sim$15, suggesting the source is
intrinsically absorbed. Table~\ref{tab_prop} summarizes quantile
results for power-law models. Only for SW3, for which $\Gamma$ is not
well-constrained, is a bremsstrahlung model used to estimate $n_H$ and
$kT$.

For the two sources with $\geq$100 counts, source spectra were grouped
to have $\geq20$ counts bin$^{-1}$ and were corrected for
background. We used {\em Sherpa} to fit power-law and
thermal-bremsstrahlung models, and account for absorption by neutral
hydrogen. Fixing $n_H$ to the appropriate values from \cite{sumi04}
gave acceptable results for both sources, in agreement with quantile
analysis; fitting for $n_H$ does not yield significant improvements
(Table~\ref{tab_specres}).

\begin{figure*} 
\centerline{\includegraphics{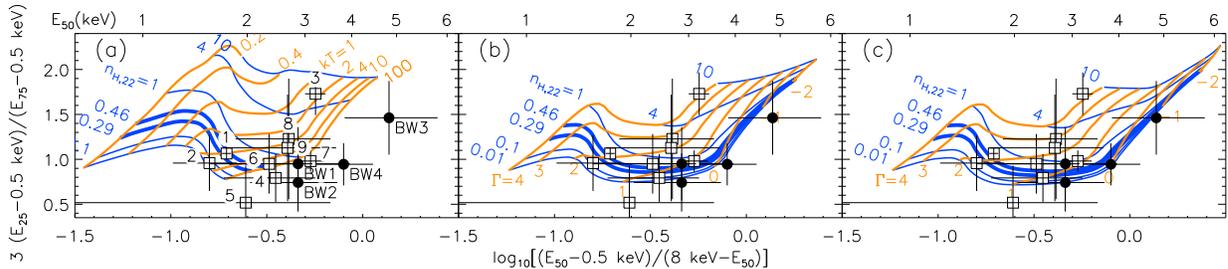}}
\caption{Quantile color-color diagrams for absorbed ($a$)
thermal-bremsstrahlung models, ($b$) power-law models and ($c$)
power-law models plus a 6.4 keV gaussian emission line (1 keV
equivalent width), showing the sources in Table~\ref{tab_prop} as
circles (\BW) and squares (\SW). Grey/orange and black/blue lines
indicate a fixed $kT$ (in keV) or $\Gamma$ and $n_{H,22}$$\equiv$$n_H$
/10$^{22}$cm$^{-2}$, respectively.  Spectral properties of a source
can be derived from its position in the grid; for example, following a
grey/orange line in ($a$) from lower left to upper right traces a
source of given $kT$ absorbed by an increasingly larger $n_H$.  Thick
lines indicate the average $n_H$ for bulge stars in \BW~and \SW. The
top axis shows the median energy $E_{50}$. {\em See electronic edition
for color figure.}
\label{fig_qccd}}
\end{figure*}

\begin{table}
\caption{Results of spectral fits (1\,$\sigma$ errors)} \label{tab_specres}
\begin{tabular}{lllllll}
\tableline
\tableline
ID         & $n_H$\tablenotemark{a} & $\Gamma$ & $\chi_\nu^{2}$/dof & $n_H$\tablenotemark{a} & $kT$ & $\chi_\nu^{2}$/dof \\
           &  & & & & (keV) \\
\tableline
SW1 & (4.3) & 2.0$\pm$0.1 & 0.52/22  & (4.3) & 3.4$^{+0.9}_{-0.6}$ & 0.60/22  \\ 
               & 2.6$^{+0.7}_{-0.6}$  & 1.7$\pm$0.1 & 0.51/21 & 2.2$^{+0.7}_{-0.6}$  &  6.0$^{+6.4}_{-1.8}$  &  0.52/21   \\ 
\tableline
SW7 & (4.1) & 0.44$^{+0.68}_{-0.73}$  & 0.014/4  & (4.1) & $<$335  & 0.27/4 \\ 
              & 8.4$^{+12}_{-6.2}$      & 0.72$^{+0.72}_{-0.77}$  &0.012/3   & 17$^{+13}_{-6.7}$       & $<$159  & 0.05/3 \\ 
\tableline
\end{tabular}
\tablenotetext{a}{in units of 10$^{21}$ cm$^{-2}$; values in
parentheses are fixed in the fit}
\end{table}

\subsection{Optical and near-Infrared Spectra} \label{sec_onirspec}

Optical spectra were obtained with the FLWO-1.5m FAST spectrograph in
2004 May and June (BW3; SW3, 4, 6, 7, 9), and the CTIO-4m Hydra
multi-object spectrograph (BW1, 3; SW3, 6, 8, 9) in 2004 June and 2005
June. FAST spectra cover 3800--7500\AA~with 3\AA~resolution while
Hydra spectra cover 4000--6800\AA\ with 4.9\AA~resolution.  Stars were
classified using templates in \cite{silvcorn92} and TiO-band strengths
\citep{kenyfern87}, see Table~\ref{tab_prop}.  H$\alpha$, and
H$\alpha$ and H$\beta$ are in emission in SW7 and 9, respectively, but
no strong high-ionization emission lines (often present in spectra of
symbiotics, see $\S$~\ref{sec_dis}) are observed. No variability is
seen for stars observed in multiple runs.

\begin{figure} 
\centerline{\includegraphics[angle=90,width=6.cm]{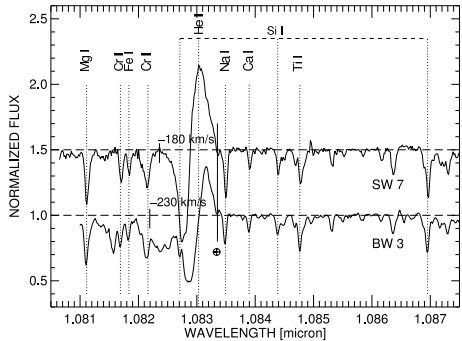}}
\caption{NIRSPEC spectra of BW3 and SW7 with identified features.
\label{fig_nirspec}}
\end{figure}

Near-infrared spectra of the He\,I~1.083\,$\mu$ line were obtained for
BW3 and SW7 on 2005 May 19 using Keck-II/NIRSPEC in high-resolution
cross-dispersed echelle mode.  Each observation consisted of a single
AB nod pair of exposures with 900\,s integrations at both slit
positions. Strong P-Cygni profiles are seen, indicative of wind
velocities of 230 (BW3) and 180\,km\,s$^{-1}$ (SW7;
Fig.\,\ref{fig_nirspec}). Similar profiles in ultraviolet lines are
found in symbiotic star systems, for example AG\,Dra
\citep{younea05}. Single M-giants do not show He\,I~1.083\,$\mu$ with
the emission strength and absorption as in BW3 and SW7 \citep{lamb87}.

\section{Discussion and conclusions} \label{sec_dis}

We explore possible explanations for the X-ray emission from these
giants. Single M-giants are rarely detected in X-rays but
\cite{huenea04} point out a few candidates that are matched securely
to X-ray sources ($L_{0.5-8 {\rm keV}}=(0.1-1.1)\times10^{31}$ erg
s$^{-1}$ for a 10$^{7}$ K coronal plasma, but $L$ is $\sim$15 times
larger for a $\Gamma=0$ power law, for $n_H=0$ cm$^{-2}$). Their
X-rays are yet unexplained, but the absence of optical emission lines
is used to argue against a symbiotic nature. In RS\,CVn binaries,
activity of a (sub)giant is increased by tidal coupling of the stellar
rotation to the orbit, but no RS\,CVns with M-giants are known. Tidal
forces in a detached binary wide enough to fit an M-giant are too weak
to spin up a main-sequence companion.  Moreover, neither single stars
nor active binaries are in general as hard as some of our sources.

Of the $\sim$200 symbiotic binaries currently known, $\sim$80\%
contain an M3--6 red giant \citep{muerschm99}. Mass transfer occurs
mostly via a wind as opposed to Roche-lobe overflow.  \cite{muerea97}
show in a ROSAT study that symbiotics are typically soft; nuclear
burning on the white dwarf produces very soft emission ($\lesssim0.4$
keV), while somewhat harder X-rays are associated with hot
($kT\approx1$ keV) shocked gas in the colliding winds of the giant and
hot star.  Note that our sample contains many hard sources
($\gtrsim50$\%, $\S$\ref{sec_xspec}) that may be similar to the
hardest sources in M{\"u}rset et al.  Emission above a few keV is so
far poorly studied for symbiotics; possible explanations include
accretion \citep[CH\,Cyg;][]{ezukea98} or shocked gas
\citep[Z\,And;][]{sokoea06}. Recently, symbiotics have been detected
also in very hard X-rays by {\em INTEGRAL} and {\em Swift}
\citep[e.g.\,RT\,Cru,][]{tuelea05}. X-ray luminosities of symbiotics
range from 10$^{29}$ to 10$^{34}$ erg s$^{-1}$, up to $\sim$10$^{37}$
erg s$^{-1}$ for the neutron-star accretor GX\,1+4.

We estimate the upper limit on the temperature of the shock-heated gas
in the colliding-wind model by requiring that the thermal energy per
gas particle does not exceed the kinetic energy. A wind velocity
$v_w=230$ km s$^{-1}$ gives $kT\lesssim0.2$ keV, lower than implied
for any of our sources by quantile analysis. Thus, if the He\,I
1.083\,$\mu$ lines in BW3 and SW7 (two of the hardest sources) trace
the conditions in the shocked winds, this model is an unlikely
explanation. To reach $kT\approx1.5$ keV as in SW3, requires
$v_w\approx660$ km s$^{-1}$. A mass-loss rate $\dot{M}=10^{-7}$
$M_\odot$ yr$^{-1}$ gives a total kinetic luminosity 1/2\,$\dot{M}
v_w^2\approx2.7 \times 10^{34}$ erg s$^{-1}$, sufficient to power
$L_{0.5-8 {\rm keV}}$=7.6$\times$($d$/8kpc)$^2\times$10$^{32}$ erg
s$^{-1}$ for SW3. Looking for such fast outflows would be one test
whether shocked winds are indeed a plausible explanation for our soft
sources.

To check whether accretion is a viable source of X-rays, we use Eq.~3
from \cite{liviwarn84} for the total accretion luminosity $L_{acc}$.
We choose masses for the giant and white dwarf of 1 and 0.6 $M_\odot$,
a white-dwarf radius of 10$^9$ cm, $\dot{M}=10^{-7}$ $M_\odot$
yr$^{-1}$, and $v_w=230$ km s$^{-1}$; the orbital speed of the white
dwarf is assumed to be large compared to the local sound speed. The
range of known orbital periods $P_b$ for symbiotics (200--5700 d)
implies 6$\times$10$^{34}\gtrsim L_{acc}\gtrsim 7\times10^{33}$ erg
s$^{-1}$, scaling linearly with $\dot{M}$. Thus it is likely that
accretion onto a white dwarf can power the X-rays: $L_{0.5-8 {\rm
keV}}$ =(2.0--110)$\times$($d$/8 kpc)$^2$ 10$^{31}$ erg s$^{-1}$
(Table~\ref{tab_prop}). Accretion onto a neutron star is less
probable: for a 1.4~$M_{\odot}$, $R=15$ km companion,
$L_{acc}\gtrsim10^{35}$ erg s$^{-1}$ for the assumed $\dot{M}$.

Assuming we observe the white dwarf through the giant's wind, we
estimate $n_H$ for the simplified case of a circular orbit and a
spherically symmetric neutral-hydrogen wind of constant velocity:
$n_H\approx\dot{M}$/(16$m_H$$v_w$$a$)~$(\cos^2{i}+\sin^2{\phi}\sin^2{i})^{-1/2}$,
with $a$ the semi-major axis, $i$ the orbital inclination, and $\phi$
the orbital phase. For the parameters and range in $P_b$ as above,
9$\times$10$^{19}\lesssim n_H \lesssim$ 2$\times$10$^{21}$ cm$^{-2}$
for $i=60^{\circ}$, below or similar to the Galactic $n_H$ for \BW~and
\SW. \cite{seaqtayl90} have found a mass-loss rate $\dot{M} \approx
10^{-8}$--10$^{-6}$ $M_\odot$ yr$^{-1}$ for red-giant symbiotics which
allows for an enhanced $n_H$, but values as high as for SW3 also
require a larger $i$. Quantile analysis suggests 6.4 keV Fe\,K
emission in our hardest sources, which may be another manifestation of
the giant's cool wind via interaction with hard X-rays; Fe\,K lines
are observed in e.g.~CH\,Cyg \citep{ezukea98} and RT\,Cru
(J.~Sokoloski 2006, private communication).

How do other properties of our sources compare to those of M-giants
and symbiotics?  Variability and mass loss, both associated with
stellar pulsations, are typical properties of single late
M-giants. \cite{glasschu02} found that many M5 and all M6 and later
giants in Baade's Window are variables. Of the 10 matches included in
\cite{woznea02}, the matches to BW1, SW7, 9, and possibly SW3, 4, 5
and 8 show semi-regular variability as commonly observed in late
M-giants (amplitudes $\lesssim2.5$ mag, time scales 20--200 d).
\cite{glasschu02} note that BW3 however shows more regular
variability. The period they find in {\em MACHO} data (135.5d) is the
same as what we find in {\em OGLE-II} data (135.2$\pm$1.5d). If this
is due to ellipsoidal variations then $P_{b}\approx270$ d.
Variability of SW2, 6 and possibly 8 is irregular with $\Delta
I\lesssim0.1$ in 4 seasons.  From mid-infrared photometry \citet{alarea01}
derive $\dot{M}\approx4\times10^{-7}$ and $4\times10^{-8}$ $M_{\odot}$
yr$^{-1}$ for the matches to BW1 and 4, respectively.  Outflow
velocities for BW3 and SW7 are high for single red giants
($v_w\approx10$--30 km s$^{-1}$) but a similar velocity has been
derived from He\,I~1.083$\mu$ for GX\,1+4 \citep{chakea98}.

Optical spectra of our matches lack strong high-ionization nebular
emission lines. The presence of such lines is one of the original
defining properties of symbiotics.  Examples of ``weakly symbiotic''
(i.e.~without strong nebular emission) systems that are hard in X-rays
are 4U\,1700+24 and 4U\,1954+319 \citep[both suspected neutron-star
systems that look like ``normal'' M-giants in the
optical,][]{maseea06} and the aforementioned {\em INTEGRAL} and {\em
Swift} sources, for which H$\alpha$ and H$\beta$ are the strongest
optical emission lines. If these are similar to our sources, the large
fraction of such systems in our sample then suggests these properties
are characteristic for a bulge sub-population and/or not identified
previously due to sensitivity limits or limited astrometric
precision. Further study of the \BW~and \SW~sources but also those of
\cite{huenea04} is therefore of interest for estimates of the total
number of Galactic symbiotics and their progeny, which may include
SNIa systems. The evolution of symbiotics depends on the orbital
period. Long-period systems become wide double white dwarfs.  For
$P_b=270$ d as for BW3, and 1 and 0.6 $M_\odot$ stars, the Roche-lobe
radius is likely $\sim94R_{\odot}$. In the models by
\cite{lejescha01}, a 1$M_\odot$ star reaches $R\approx110R_{\odot}$ at
the tip of the red-giant branch and expands even further as an
asymptotic giant. If BW3 is a binary, it may start unstable mass
transfer leaving a close double white dwarf, which could evolve to a
SN\,Ia.

We use the results by \citet{laycea05} to constrain the number of
symbiotics among the initial sample of 1453 still unclassified faint,
hard X-ray sources in the 10\arcmin~$\times$ 10\arcmin~region around
the GC \citep{munoea03}. For $A_{K,GC}\approx3.4$ and $M_K\lesssim-4$,
$K\lesssim13.9$ for M-giants at the GC. Laycock et al.~find that
$\lesssim7$\% (90\% confidence) of the 110 hard sources within
1\arcmin~from the GC can have counterparts with $K\leq14$, while this
is true for $\leq4$\% of the 1343 hard sources at larger
offsets. These likely include high-mass X-ray binaries with B0 V
primaries but also symbiotics.

\begin{acknowledgements}
We thank J.~Sokoloski for discussions and comments on the
manuscript. This work is supported by Chandra grants GO3-4033A and
AR6-7010X. We are grateful for the use of the Keck Observatory on
Mauna Kea.
\end{acknowledgements}


\begin{thebibliography}{37}
\expandafter\ifx\csname natexlab\endcsname\relax\def\natexlab#1{#1}\fi

\bibitem[{{Alard} {et~al.}(2001){Alard}, {Blommaert}, {Cesarsky}, {Epchtein},
  {Felli}, {Fouque}, {Ganesh}, {Genzel}, {Gilmore}, {Glass}, {Habing}, {Omont},
  {Perault}, {Price}, {Robin}, {Schultheis}, {Simon}, {van Loon}, {Alcock},
  {Allsman}, {Alves}, {Axelrod}, {Becker}, {Bennett}, {Cook}, {Drake},
  {Freeman}, {Geha}, {Griest}, {Lehner}, {Marshall}, {Minniti}, {Nelson},
  {Peterson}, {Popowski}, {Pratt}, {Quinn}, {Sutherland}, {Tomaney},
  {Vandehei}, \& {Welch}}]{alarea01}
{Alard}, C. {et~al.} 2001, \apj, 552, 289

\bibitem[{{Blanco}(1986)}]{blan86}
{Blanco}, V. 1986, \aj, 91, 290

\bibitem[{{Blanco} {et~al.}(1984){Blanco}, {McCarthy}, \&
  {Blanco}}]{blanmccaea84}
{Blanco}, V., {McCarthy}, M., \& {Blanco}, B. 1984, \aj, 89, 636

\bibitem[{{Chakrabarty} {et~al.}(1998){Chakrabarty}, {van Kerkwijk}, \&
  {Larkin}}]{chakea98}
{Chakrabarty}, D., {van Kerkwijk}, M.~H., \& {Larkin}, J.~E. 1998, \apjl, 497,
  L39+

\bibitem[{{Ezuka} {et~al.}(1998){Ezuka}, {Ishida}, \& {Makino}}]{ezukea98}
{Ezuka}, H., {Ishida}, M., \& {Makino}, F. 1998, \apj, 499, 388

\bibitem[{Girardi(2006)}]{gira06}
Girardi, L. 2006, http://pleiadi.pd.astro.it

\bibitem[{{Glass} \& {Schultheis}(2002)}]{glasschu02}
{Glass}, I.~S., \& {Schultheis}, M. 2002, \mnras, 337, 519

\bibitem[{Grindlay {et~al.}(2005)Grindlay, Hong, Zhao, Laycock, van~den Berg,
  Koenig, , Schlegel, Cohn, Lugger, \& Rogel}]{grinhongea05}
Grindlay, J. {et~al.} 2005, ApJ, 635, 920

\bibitem[{{Hong} {et~al.}(2004){Hong}, {Schlegel}, \&
  {Grindlay}}]{hongschlea04}
{Hong}, J., {Schlegel}, E.~M., \& {Grindlay}, J.~E. 2004, \apj, 614, 508

\bibitem[{Hong {et~al.}(2005)Hong, van~den Berg, Schlegel, Grindlay, Koenig,
  Laycock, \& Zhao}]{hongvandea05}
Hong, J., van~den Berg, M., Schlegel, E., Grindlay, J., Koenig, X., Laycock,
  S., \& Zhao, P. 2005, ApJ, 635, 907

\bibitem[{{Houdashelt} {et~al.}(2000){Houdashelt}, {Bell}, {Sweigart}, \&
  {Wing}}]{houdbellea00}
{Houdashelt}, M., {Bell}, R., {Sweigart}, A., \& {Wing}, R.~F. 2000, \aj, 119,
  1424

\bibitem[{{H{\"u}nsch} {et~al.}(2004){H{\"u}nsch}, {Konstantinova-Antova},
  {Schmitt}, {Schr{\"o}der}, {Kolev}, {de Medeiros}, {L{\`e}bre}, \&
  {Udry}}]{huenea04}
{H{\"u}nsch}, M., {Konstantinova-Antova}, R., {Schmitt}, J., {Schr{\"o}der},
  K.-P., {Kolev}, D., {de Medeiros}, J.-R., {L{\`e}bre}, A., \& {Udry}, S.
  2004, in IAU Symp., 223

\bibitem[{{H{\"u}nsch} {et~al.}(1998){H{\"unsch}}, {Schmitt}, {Schr{\"o}der}, \&
  {Zickgraf}}]{huenea98}
{H{\"u}nsch}, M., {Schmitt}, J., {Schr{\"o}der}, K.-P., \& {Zickgraf}, F.-J. 1998,
\aap, 330, 225

\bibitem[{{Kenyon} \& {Fernandez-Castro}(1987)}]{kenyfern87}
{Kenyon}, S.~J., \& {Fernandez-Castro}, T. 1987, \aj, 93, 938

\bibitem[{{Lambert}(1987)}]{lamb87}
{Lambert}, D.~L. 1987, \apjs, 65, 255

\bibitem[{{Laycock} {et~al.}(2005){Laycock}, {Grindlay}, {van den Berg},
  {Zhao}, {Hong}, {Koenig}, {Schlegel}, \& {Persson}}]{laycea05}
{Laycock}, S., {Grindlay}, J., {van den Berg}, M., {Zhao}, P., {Hong}, J.,
  {Koenig}, X., {Schlegel}, E., \& {Persson}, S.~E. 2005, \apjl, 634, L53

\bibitem[{{Lejeune} \& {Schaerer}(2001)}]{lejescha01}
{Lejeune}, T., \& {Schaerer}, D. 2001, \aap, 366, 538

\bibitem[{{Livio} \& {Warner}(1984)}]{liviwarn84}
{Livio}, M., \& {Warner}, B. 1984, The Observatory, 104, 152

\bibitem[{{Masetti} {et~al.}(2006){Masetti}, {Orlandini}, {Palazzi},
 {Amati}, \& {Frontera}}]{maseea06}
 {Masetti}, N., {Orlandini}, M., {Palazzi}, E., {Amati}, L., \& {Frontera}, F., 
 \aap, in press (astro-ph/0603227)

\bibitem[{{Muno} {et~al.}(2003){Muno}, {Baganoff}, {Bautz}, {Brandt}, {Broos},
  {Feigelson}, {Garmire}, {Morris}, {Ricker}, \& {Townsley}}]{munoea03}
{Muno}, M.~P. {et~al.} 2003, \apj, 589, 225

\bibitem[{{M{\"u}rset} \& {Schmid}(1999)}]{muerschm99}
{M{\"u}rset}, U., \& {Schmid}, H.~M. 1999, \aaps, 137, 473

\bibitem[{{M{\"u}rset} {et~al.}(1997){M{\"u}rset}, {Wolff}, \&
  {Jordan}}]{muerea97}
{M{\"u}rset}, U., {Wolff}, B., \& {Jordan}, S. 1997, \aap, 319, 201

\bibitem[{Schmitt \& Liefke(2004)}]{schmlief04}
Schmitt, J. H. M.~M., \& Liefke, C. 2004, A\&A, 417, 651

\bibitem[{{Seaquist} \& {Taylor}(1990)}]{seaqtayl90}
{Seaquist}, E.~R., \& {Taylor}, A.~R. 1990, \apj, 349, 313

\bibitem[{{Silva} \& {Cornell}(1992)}]{silvcorn92}
{Silva}, D.~R., \& {Cornell}, M.~E. 1992, \apjs, 81, 865

\bibitem[{{Sokoloski} {et~al.}(2006){Sokoloski}, {Kenyon}, {Espey}, {Keyes},
  {McCandliss}, {Kong}, {Aufdenberg}, {Filippenko}, {Li}, {Brocksopp},
  {Kaiser}, {Charles}, {Rupen}, \& {Stone}}]{sokoea06}
{Sokoloski}, J.~L. {et~al.} 2006, \apj, 636, 1002

\bibitem[{Stanek(1998)}]{stan98}
Stanek, K.~Z. 1998, astro-ph/9802307

\bibitem[{{Sumi}(2004)}]{sumi04}
{Sumi}, T. 2004, \mnras, 349, 193

\bibitem[{{Tueller} {et~al.}(2005){Tueller}, {Gehrels},
  {Mushotzky}, {Markwardt}, {Kennea}, {Burrows}, {Mukai}, \&
  {Sokoloski}}]{tuelea05}
{Tueller}, J., {Gehrels}, N., {Mushotzky}, R., {Markwardt}, C., {Kennea}, J.,
  {Burrows}, D., {Mukai}, K., \& {Sokoloski}, J. 2005, The
  Astronomer's Telegram, 591, 1

\bibitem[{{Udalski} {et~al.}(2002){Udalski}, {Szymanski}, {Kubiak},
  {Pietrzynski}, {Soszynski}, {Wozniak}, {Zebrun}, {Szewczyk}, \&
  {Wyrzykowski}}]{udalszymea02}
{Udalski}, A. {et~al.} 2002, Acta Astronomica, 52, 217 (U02)

\bibitem[{{Wozniak} {et~al.}(2002){Wozniak}, {Udalski}, {Szymanski}, {Kubiak},
  {Pietrzynski}, {Soszynski}, \& {Zebrun}}]{woznea02}
{Wozniak}, P.~R., {Udalski}, A., {Szymanski}, M., {Kubiak}, M., {Pietrzynski},
  G., {Soszynski}, I., \& {Zebrun}, K. 2002, Acta Astronomica, 52, 129

\bibitem[{{Young} {et~al.}(2005){Young}, {Dupree}, {Espey}, {Kenyon}, \&
  {Ake}}]{younea05}
{Young}, P.~R., {Dupree}, A.~K., {Espey}, B.~R., {Kenyon}, S.~J., \& {Ake},
  T.~B. 2005, \apj, 618, 891

\bibitem[{Zhao {et~al.}(2005)Zhao, Grindlay, Hong, Laycock, Koenig, Schlegel,
  \& van~den Berg}]{zhaogrindea05}
Zhao, P., Grindlay, J., Hong, J., Laycock, S., Koenig, X., Schlegel, E., \&
  van~den Berg, M. 2005, ApJS, 161, 429

\bibitem[{{Zheng} {et~al.}(2001){Zheng}, {Flynn}, {Gould}, {Bahcall}, \&
  {Salim}}]{zhenea01}
{Zheng}, Z., {Flynn}, C., {Gould}, A., {Bahcall}, J.~N., \& {Salim}, S. 2001,
  \apj, 555, 393

\end{thebibliography}
\end{document}